\documentclass[rmp,twocolumn,superbib]{revtex4}%
\usepackage{amsfonts}
\usepackage{times}
\usepackage{amsmath}
\usepackage{amssymb}
\usepackage{graphicx}%
\begin{document}
\title{Precessionless spin transport wire confined in quasi-two-dimensional electron systems}
\author{Ming-Hao Liu}
\email{d92222010@ntu.edu.tw}
\affiliation{Department of Physics, National Taiwan University, Taipei, Taiwan}
\author{Son-Hsien Chen}
\affiliation{Department of Physics, National Taiwan University, Taipei, Taiwan}
\author{Ching-Ray Chang}
\affiliation{Department of Physics, National Taiwan University, Taipei, Taiwan}

\begin{abstract}
We demonstrate that in an inversion-asymmetric two-dimensional electron system
(2DES) with both Rashba and Dresselhaus spin-orbit couplings taken into
account, certain transport directions on which no spin precession occurs can
be found when the injected spin is properly polarized. By analyzing the
expectation value of spin with respect to the injected electron state on each
space point in the 2DES, we further show that the adjacent regions with
technically reachable widths along these directions exhibit nearly conserved
spin. Hence a possible application in semiconductor spintronics, namely,
precessionless spin transport wire is proposed.

\end{abstract}
\date[Presented on ]{1 November 2005}
\maketitle

Crystallographic-direction-dependenc$^{\text{\cite{LusakowskiA2003,TingDZY}}}$
of the Datta-Das transistor$^{\text{\cite{SFET}}}$ and the spin orientation in
an inversion-asymmetric two-dimensional electron system
(2DES)$^{\text{\cite{Winkler,RDSP}}}$ have been investigated recently. In the
ballistic regime, electron spins are expected to precess when propagating in
the 2DES, where electrons encounter spin-orbit (SO) couplings due to space
inversion asymmetry, including structure and bulk. The former, structure
inversion asymmetry (SIA),$^{\text{\cite{Rashba term,Rashba term 2}}}$
generates the well-known Rashba spin-orbit (RSO) coupling with strength
$\alpha$ being gate-voltage tunable$^{\text{\cite{Nitta}}}$ while the latter,
bulk inversion asymmetry (BIA), induces the Dresselhaus spin-orbit (DSO)
coupling,$^{\text{\cite{Dresselhaus term,LommerG1985,Dyakonov}}}$ with
strength $\beta$ being material specific. When an electron with specific spin
is ideally injected into the 2DES, the electron state is superposed by the two
spin-dependent eigenstates with a phase difference between. As determining the
electron spin away from the injection point, the expectation value, in
general, differs from the injected one due to interference between the two
superposing components, and the spin precession thus occurs.%
\begin{figure}
[pb]
\begin{center}
\includegraphics[
height=3.9686in,
width=2.015in
]%
{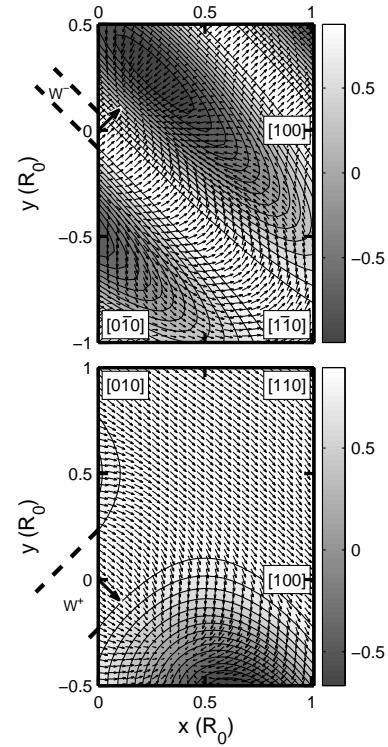}%
\caption{Spin vectors of the $\pi/4$-oriented (upper panel) and $-\pi
/4$-oriented (lower panel) injected spin [indicated by the bold arrow on
$\left(  0,0\right)  $] in the 2DES channel where both RSO and DSO are
nonvanishing with their coupling strength ratio $\alpha/\beta=2.15$. The
gray-scale background is determined by the projections between local spin
vectors and the injected spin. The compact unit $R_{0}$ given by Eq. (1) is
the precession period length along [100] or [010] axes.}%
\label{STW}%
\end{center}
\end{figure}

Recently this spin precession due to individually the RSO and DSO couplings
has been mathematically investigated and pictorially
introduced.$^{\text{\cite{RDSP}}}$ Implications indicated therein had
demonstrated the uniqueness of the four crystallographic directions $\pm
$[1$\pm$10] for a [001]-grown zinc-blend-based 2DES. The spatial behaviors on
these four directions are shown to be invariant, regardless of the coupling
ratio $\alpha/\beta$, since the effective magnetic fields generated by RSO and
DSO are either parallel or antiparallel to each other on these axes. More
specifically, the total effective magnetic field directions are always
perpendicular to the electron wave vectors when propagating along $\pm$[1$\pm
$10]. Possible applications utilizing these four axes are thus envisioned. In
this paper we apply the theoretical method constructed in Ref.
$^{\text{\cite{RDSP}}}$ to demonstrate that in the 2DES with nonvanishing RSO
and DSO couplings, zero spin precession (ZSP) axes, i.e., axes on which no
spin precession occurs, can be found when the injected spin is properly
oriented. By analyzing the spatially varying spin vectors, i.e., spin
expectation values done with respect to the injected spin states, we further
show that the adjacent regions with technically reachable widths along these
directions own nearly conserved spin. Hence a possible application in
semiconductor spintronics, namely, precessionless spin transport wire (STW) is proposed.

We begin by giving two examples of the STW confined in a [001]-grown 2DES. In
the upper panel of Fig. \ref{STW}, we inject a $\pi/4$-polarized spin
(polarization angle relative to [100]) on the 2DES where the RSO and DSO
coupling ratio is set $\alpha/\beta=2.15$.$^{\text{\cite{GanichevSD2004}}}$
Shading the channel with projections between local spin vectors and the
injected spin, a narrow precessionless path along [1\={1}0] is clearly
observed. In this case the ZSP axis is definitely [1\={1}0] since we inject a
spin parallel to the effective magnetic field thereon. Note that the
longitudinal and transverse positions of the 2DES channel is labeled in units
of $R_{0}$ defined by
\begin{equation}
R_{0}\equiv\frac{\pi\hbar^{2}}{m^{\ast}\sqrt{\alpha^{2}+\beta^{2}}}\text{.}%
\end{equation}
Taking the DSO coupling strength as $\beta=0.1%
\operatorname{eV}%
\operatorname{\text{\AA}}%
$, which is a typical value estimated for a $50%
\operatorname{\text{\AA}}%
$ thick quantum well in III-V semiconductors, and the effective mass $m^{\ast
}=0.023m_{0}$, corresponding to the InAs type quantum wells, $R_{0}$ is about
$0.44%
\operatorname{\mu m}%
$.

When injecting a $-\pi/4$-polarized spin, a wider precessionless region is
observed (lower panel in Fig. \ref{STW}). A basic difference between these two
cases is that the anisotropic total SO coupling strength $\gamma\left(
\phi\right)  =\sqrt{\alpha^{2}+\beta^{2}+2\alpha\beta\sin\left(  2\phi\right)
}$, where $\phi$ is the wave vector direction relative to [100], becomes
strongest, i.e., $\gamma\left(  \pi/4\right)  =\left\vert \alpha
+\beta\right\vert $ along [110] while weakest, i.e., $\gamma\left(
-\pi/4\right)  =\left\vert \alpha-\beta\right\vert $ along [1\={1}0].
Therefore, the injected spin encounters weaker (stronger) spin precession when
propagating perpendicular to the wire in the [110] ([1\={1}0]) case, leading
to a wider (narrower) precessionless region.%
\begin{figure}
[ptb]
\begin{center}
\includegraphics[
height=1.977in,
width=2.143in
]%
{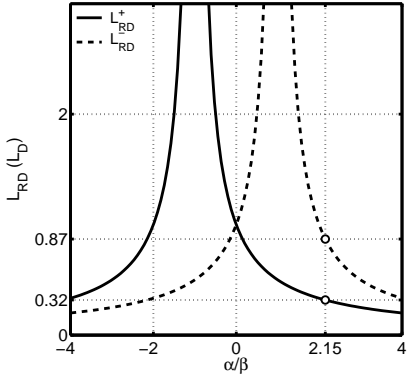}%
\caption{The half period distances $L_{RD}^{\pm}$ as a function of the
coupling ratio $\alpha/\beta$.}%
\label{WidthOfSTW}%
\end{center}
\end{figure}

Despite the direction dependence, the coupling ratio may also influence the
wire width. Defining the half precession length%
\begin{equation}
L_{RD}=\frac{\pi\hbar^{2}}{2m^{\ast}\sqrt{\alpha^{2}+\beta^{2}+2\alpha
\beta\sin2\phi}}\text{,}%
\end{equation}
the wire widths can be approximated by $W^{\pm}\approx L_{RD}^{\mp}/2$ with%
\begin{equation}
L_{RD}^{\pm}\equiv L_{RD}\left(  \phi=\pm\frac{\pi}{4}\right)  =\frac{L_{D}%
}{\left\vert \alpha/\beta\pm1\right\vert }%
\end{equation}
where $L_{D}\equiv\pi\hbar^{2}/\left(  2m^{\ast}\beta\right)  $. Note that the
singular points for $L_{RD}^{\pm}$ at $\alpha/\beta=\mp1$ correspond to the
unique "cancellation" of the Rashba and Dresselhaus terms, removing the
$k$-dependence of the eigenspinors.$^{\text{\cite{NBSFET}}}$ We plot $L_{RD}$
along $\phi=\pm\pi/4$ as a function of $\alpha/\beta$ in Fig. \ref{WidthOfSTW}%
, where $L_{RD}^{\pm}\left(  \alpha/\beta=2.15\right)  $ are also marked out,
showing $L_{RD}^{-}\left(  \alpha/\beta=2.15\right)  >L_{RD}^{+}\left(
\alpha/\beta=2.15\right)  $ so that $W^{+}>W^{-}$ in Fig. \ref{STW} is
definitely true. An estimate for the DSP half precession length $L_{D}$ can
also be made by substituting $\beta\approx0.1%
\operatorname{eV}%
\operatorname{\text{\AA}}%
^{\text{\cite{NBSFET}}}$ and $m^{\ast}/m_{0}=0.023$.$^{\text{\cite{Knap}}}$
This yields $L_{D}\approx0.52%
\operatorname{\mu m}%
$, leading to $W^{+}\approx\allowbreak226%
\operatorname{nm}%
$ and $W^{-}\approx83%
\operatorname{nm}%
$ which are technically reachable scales in current semiconductor industry. As
can be seen from Fig. \ref{WidthOfSTW}, the wire width can even be
considerably broaden by properly adjusting the coupling ratio $\alpha/\beta$.%
\begin{figure}
[pb]
\begin{center}
\includegraphics[
height=2.1837in,
width=3.4151in
]%
{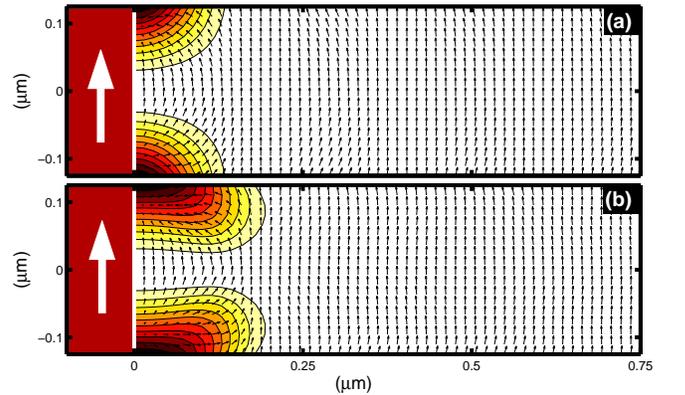}%
\caption{(Color online) Spin vectors inside a [110] channel with size
$0.25\times0.75\operatorname{\mu m}^{2}$, using a wide and perfectly polarized
source contact for spin injection, (a) in the absence and (b) in the presence
of channel boundary effects.}%
\label{FIG3}%
\end{center}
\end{figure}

Next we take into account the size effect of the spin injection contact and
the possible influence due to the channel boundary.$^{\text{\cite{MHL}}}$ In
the former consideration, the electron spin state on the space points inside
the wire will be a superposition of every contribution from the spin injection
contact, while in the latter modification, additional contribution from the
state kets reflected from the two sides of the wire needs also to be summed.
We consider a $0.25%
\operatorname{\mu m}%
\times0.75%
\operatorname{\mu m}%
$ [110] channel, using a full side contact with perfect polarization of
magnetization. The coupling parameters are set as $\alpha=0.3%
\operatorname{eV}%
\operatorname{\text{\AA}}%
$ and $\beta=0.09%
\operatorname{eV}%
\operatorname{\text{\AA}}%
$. As shown in Fig. \ref{FIG3}, the anisotropy of the spin vectors occurs only
in regions near the source contact, whether the boundary effect is taken into
account, and hence does not influence the collection of the spin signal at the
end of the channel. When comparing with the point injection case, these
additional consideration does not raise considerable
change,$^{\text{\cite{MHL}}}$ in particular, for channel directions with
strong SO strength such as [110]. Therefore, the analysis for point injection
cases introduced previously may work well enough.

It is worthy of mentioning that the\emph{ }effect of the boundary scattering
on the dynamics of propagating spin in clean systems has recently received
increased attention for both regular and chaotic walls.$^{\text{\cite{Chang CH
2004,Zaitsev 2005,Nikolic 2005}}}$ However, such relaxation, namely the
D'yakonov-Perel' (DP) spin relaxation, is suppressed in narrow
wires$^{\text{\cite{Malshukov2000,Kiselev}}}$ and essentially in strictly
single channel quantum transport.$^{\text{\cite{Nikolic 2005,DPrelax}}}$ Thus
the DP spin relaxation, neglected in our calculation, does not destroy our
analysis on the STW. In fact, it's relaxation in magnitude of the spin vectors
appears insignificant for spin transport within the order of micrometer from
Monte Carlo simulation.$^{\text{\cite{DPrelax}}}$

We now turn back to the point injection case and consider the same geometry
but in three different channel directions [100], [110], and [1-10]. Defining
the averaged spin signal collected at the end of the wire as the sum of the
components along the drain contact magnetization direction (assumed transverse
here) of all the spin vectors calculated thereat divided by the wire width
($0.25%
\operatorname{\mu m}%
$ in this case), we examine the gate-voltage dependence of these three wires,
i.e., we examine the collected spin signal by varying the Rashba coupling
strength $\alpha$.
\begin{figure}
[ptb]
\begin{center}
\includegraphics[
height=1.9804in,
width=2.5253in
]%
{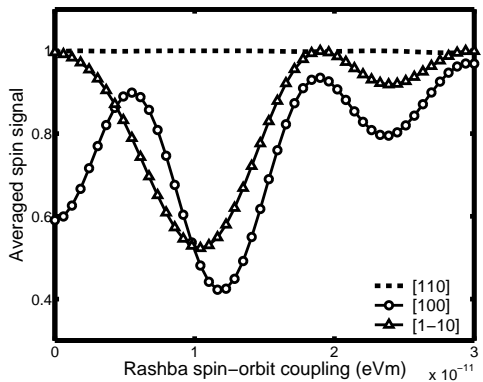}%
\caption{The averaged spin signal (in units of $\hbar/2$) collected at the end
of the $0.25\operatorname{\mu m}\times0.75\operatorname{\mu m}$ wire for
[100], [110], and [1-10] cases. The calculation is done within the single
injection method.}%
\label{FIG4}%
\end{center}
\end{figure}
As can be seen in Fig. \ref{FIG4}, the spin signal oscillates with the
increasing Rashba field for the [100] wire, which is obviously not
precessionless. For the two promising cases [1$\pm$10], only the [110]
exhibits precessionless behavior. This is because the width of the
precessionless region along [1-10] is much narrower than the [110] case
(mentioned previously) and thus the spin signal is averaged down when summing
the spin vectors away from the [1-10] axis. With the above analysis, we also
suggest the corresponding experimental setup to inspect the applicability of
our precessionless STW.

In conclusion, we have demonstrated that the electron spin may be maintainably
and perpendicularly transported along $\pm$[1$\pm$10] in a zinc-blend-based
[001]-grown 2DES within a reasonable width. Such a significant property of the
inversion-asymmetric 2DES can be designed to be a precessionless spin
transport wire or channel with widths depending on the coupling ratio of
Rashba and Dresselhaus terms. In general, one can determine the effective
field direction for a specific wave vector by measuring $\alpha/\beta$
precisely, inject spins polarized either parallel or antiparallel to the
corresponding field, and then obtain precessionless spin transport. In such
cases the spin polarization is even not necessarily perpendicular to its
propagation. However, we emphasize that our results reveal that the spin
transport along the four axes $\pm$[1$\pm$10] are always precessionless when
injecting perpendicularly polarized spins, regardless of the magnitudes of the
RSO and DSO coupling strengths. In particular, [110] is recommended in the
case of $\alpha\beta>0$ for having wider precessionless region and being more
robust against the influence due to finite-size spin injection and the channel
boundary reflections.$^{\text{\cite{MHL}}}$ Such precessionless STW, if
successfully fabricated, may be a basic component in the future spin-related
devices. For example, the precessionless STW may serve as the lead between the
spin source contact and the semiconductor-based transport channel to lower the
loss of polarization of spin injection. Hence a modified version of the
Datta-Das spin-field-effect transistor can also be proposed.

This work is supported by the Republic of China National Science Council Grant
No. 94-2112-M-002-004.


\begin{thebibliography}{99}                                                                                               %


\bibitem {LusakowskiA2003}A \L usakowski, J. Wr\'{o}bel, and T. Dietl, Phys.
Rev. B \textbf{68}, R081201 (2003).

\bibitem {TingDZY}David Z.-Y. Ting and Xavier Cartoix\`{a}, Phys. Rev. B,
\textbf{68 }235320 (2003).

\bibitem {SFET}S. Datta and B. Das, Appl. Phys. Lett. \textbf{56}, 665 (1990).

\bibitem {Winkler}R. Winkler, Phys. Rev. B \textbf{69}, 045317 (2004).

\bibitem {RDSP}Ming-Hao Liu, Ching-Ray Chang, and Son-Hsien Chen, Phys. Rev. B
\textbf{71}, 153305 (2005).

\bibitem {Rashba term}E. I. Rashba, Sov. Phys. Solid State \textbf{2}, 1109 (1960);

\bibitem {Rashba term 2}Yu. A. Bychkov and E. I. Rashba, JETP Lett.
\textbf{39}, 78 (1984).

\bibitem {Nitta}J. Nitta, T. Akazaki, H. Takayanagi, and T. Enoki, Phys. Rev.
Lett. \textbf{78}, 1335 (1997).

\bibitem {Dresselhaus term}G. Dresselhaus, Phys. Rev. \textbf{100}, 580 (1955).

\bibitem {LommerG1985}G. Lommer, F. Malcher, and U. R\"{o}ssler, Phys. Rev. B
\textbf{32}, 6965 (1985);

\bibitem {Dyakonov}M. I. D'yakonov and V. Y. Kachorovskii, Sov. Phys.
Semicond. \textbf{20}, 110 (1986).

\bibitem {GanichevSD2004}S. D. Ganichev, V. V. Bel'kov, L. E. Golub, E. L.
Ivchenko, P. Schneider, S. Giglberger, J. Eroms, J. De Boeck, G. Borghs, W.
Wegscheider, D. Weiss, and W. Prettl, Phys. Rev. Lett. \textbf{92}, 256601 (2004).

\bibitem {NBSFET}John Schliemann, J. Carlos Egues, and Daniel Loss, Phys. Rev.
Lett. \textbf{90}, 146801 (2003).

\bibitem {Knap}W. Knap, C. Skierbiszewski, A. Zduniak, E. Litwin-Staszewska,
D. Bertho, F. Kobbi, J. L. Robert, G. E. Pikus, F. G. Pikus, S. V. Iordanskii,
V. Mosser, K. Zekentes, and Yu. B. Lyanda-Geller, Phys. Rev. B \textbf{53},
3912 (1996)

\bibitem {MHL}Ming-Hao Liu and Ching-Ray Chang, cond-mat/0507495.

\bibitem {Chang CH 2004}Cheng-Hung Chang, A. G. Mal'shukov, and K. A. Chao,
Phys. Rev. B \textbf{70}, 245309 (2004).

\bibitem {Zaitsev 2005}Oleg Zaitsev, Diego Frustaglia, and Klaus Richter,
Phys. Rev. Lett. \textbf{94}, 026809 (2005).

\bibitem {Nikolic 2005}Branislav K. Nikoli\'{c} and Satofumi Souma, Phys. Rev.
B \textbf{71}, 195328 (2005).

\bibitem {Malshukov2000}A. G. Mal'shukov and K. A. Chao, Phys. Rev. B
\textbf{61}, R2413 (2000).

\bibitem {Kiselev}A. A. Kiselev and K. W. Kim, Phys. Rev. B \textbf{61}, 13115 (2000).

\bibitem {DPrelax}Sandipan Pramanik, Supriyo Bandyopadhyay, and Marc Cahay, cond-mat/0403021.
\end{thebibliography}
\end{document}